\documentclass[a4paper,11pt]{article}
\usepackage{pos}

\title{A Novel Approach towards the Search for Gamma-ray Emission from the Northern \textit{Fermi} Bubble with HAWC}
 \ShortTitle{Nothern \textit{Fermi} Bubble with HAWC}

\author*[a]{Pooja Surajbali}

\affiliation[a]{Max Planck Institut f\"ur Kernphysik,\\
  Saupfercheckweg 1, Heidelberg, Germany}

\forColl{HAWC} 

\emailAdd{pooja@mpi-hd.mpg.de}

\abstract{The {\it Fermi} bubbles are structures observed in gamma rays at GeV energies, emanating from the central region of our galaxy and extending up to 8.5 kpc above and below the galactic plane. While initial studies showed a flat brightness across the entire structure, more recent work found a brightening at the base. We perform a template-based search for TeV signals from the northern \textit{Fermi} bubble and just from the base of it in data from the High Altitude Water Cherenkov (HAWC) gamma-ray observatory. We employ a profile likelihood approach to calculate the significance and flux from the search regions. With no significant signal from the northern \textit{Fermi} bubble and its base, we report new upper limits on the integral flux at $95\%$ confidence level. Our integral flux upper limits for the northern \textit{Fermi} bubble are more constraining than the previous limits reported by HAWC. Moreover, we present, for the first time, TeV limits pertaining to the base of the bubble which constitutes a more fair comparison to \textit{Fermi} Large Area Telescope data points close to this particular region.}

\FullConference{37$^{\rm{th}}$ International Cosmic Ray Conference (ICRC 2021)\\
		July 12th -- 23rd, 2021\\
		Online -- Berlin, Germany}

\begin{document}
\maketitle

\section{Introduction}
\label{intro}
The High Altitude Water Cherenkov (HAWC) gamma-ray observatory is an air shower particle detector, located at an altitude of $4\,100$ metres a.s.l., close to Pico de Orizaba, Mexico, with the coordinates being $97.3 \degree$W and $19.0 \degree$N. HAWC was inaugurated in March 2015 and consists of an array of water tanks that function as water Cherenkov detectors designed to operate in the gamma-ray energy range of $300$ GeV to $100$ TeV. It continuously surveys the sky overhead and is ideal for studying large-scale structures. 

\medskip
Some of the largest gamma-ray structures in our Milky Way galaxy are the \textit{Fermi} bubbles. These are gamma-ray structures that appear to be emanating from the central part of the galaxy. They are tremendous in size, extending up to $55\degree$ ($\equiv 8$ kpc) above and below the Galactic plane. Although initial reports have claimed a sharp cut-off in the spectrum of the \textit{Fermi} bubbles and a flat projected brightness distribution \cite{su2010giant, Ackermann2014Spectrum}, there is a gradient in the surface brightness profile of the bubbles \cite{Surajbali2020Observing}. A recent study \cite{Herold2019Hard} provides evidence of harder and brighter gamma-ray emission coming from the base of the bubble which fits a power-law spectrum without cut-off.

\medskip
A search of very high energy gamma-ray signal from the northern \textit{Fermi} bubble with HAWC was presented in 2017 \cite{HAWC2017search}. With only $290$ days of data, no significant excess was observed from the region corresponding to the northern \textit{Fermi} bubble and differential upper limits above $1$ TeV were computed. We perform another search in a similar region with $800$ days of HAWC data, new sets of gamma/hadron cuts, a more appropriate background model, and a different approach for computing significance \cite{Surajbali2020Observing}. The \textit{Fermi} bubble excess is expected to be faint and distributed over an extended region. Therefore, we implement both a uniform and a weighted spatial template to perform a signal search from the northern \textit{Fermi} bubble and just from its base. The uniform template for the northern \textit{Fermi} bubble, henceforth NFB, was obtained from \texttt{Fermi Tools} and truncated at latitude $10\degree$, as illustrated in Figure~\ref{wedge_des}. In the following subsections, we describe how we derive the uniform template for the base for the northern \textit{Fermi} bubble and the weighted spatial template for both regions.

\subsection{Deriving Uniform Spatial Template for the Base of the Northern \textit{Fermi} Bubble}
\label{UniformTemp}

\begin{figure}[htbp]
\centering
\includegraphics[width=0.7\hsize]{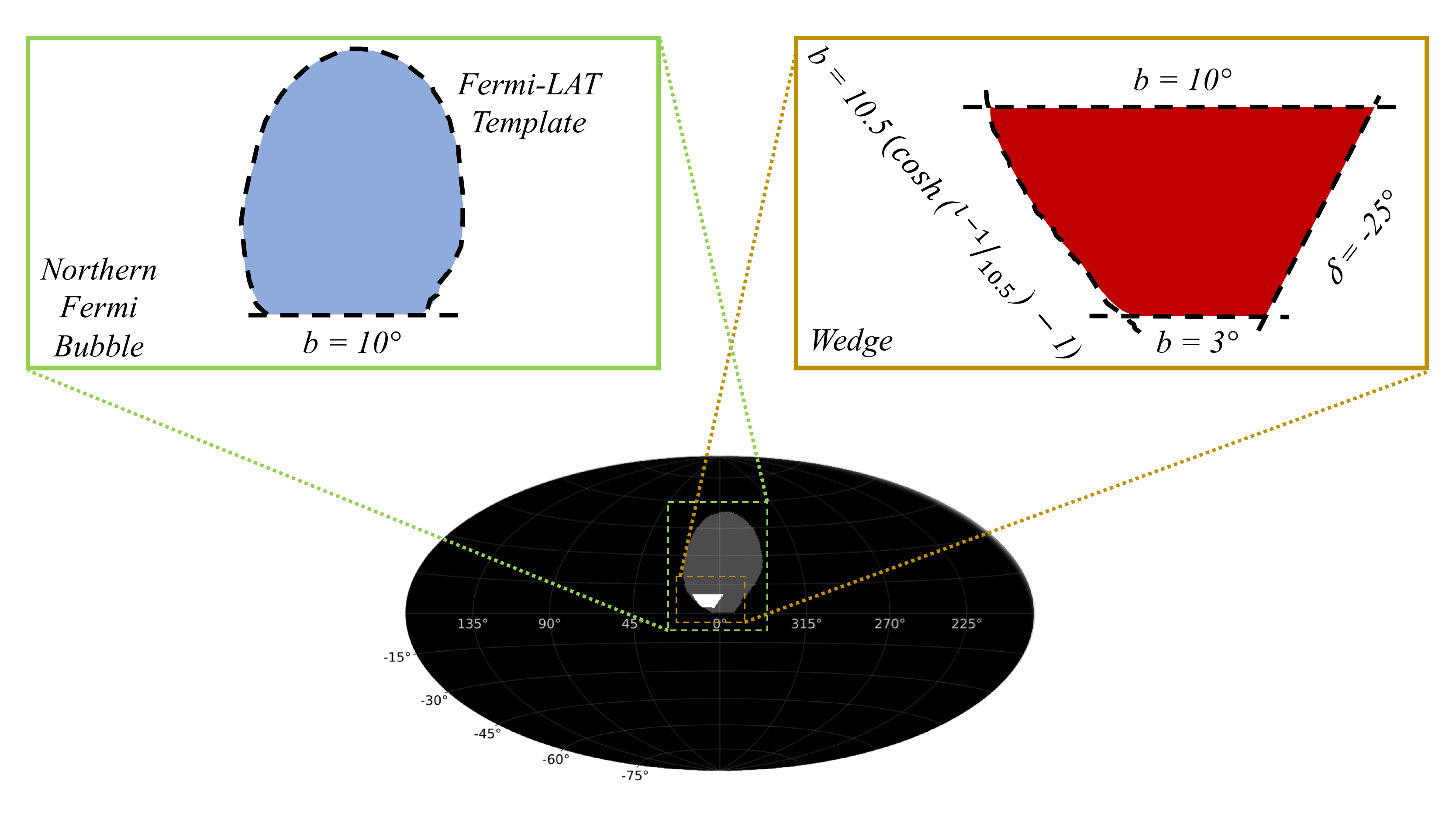}
\caption{Description of search regions, illustrating the northern \textit{Fermi} bubble (NFB) and the wedge at the base of the northern \textit{Fermi} bubble.}
\label{wedge_des}
\end{figure}

The sensitivity of HAWC is dependent on declination, we therefore select a region which most likely contains the signal for the prominent part of the emission. We define a four-sided search region, henceforth referred to as the \textit{wedge}, as illustrated in Figure~\ref{wedge_des}, with the following boundaries:
\begin{enumerate}
\item $b = 10\degree$: The region below this line is typically referred to as the base of the \textit{Fermi} bubble;

\item $b = 3\degree$: Limit to avoid most of Galactic plane emission;

\item $\delta = -25\degree$: Declination limit for HAWC field of view;

\item $b = 10.5 \big(\mathrm{cosh}(\frac{l - 1}{10.5}) - 1 \big)$: Bubble boundary obtained by comparing the \textit{Fermi} Large Area Telescope (LAT) data to the X-ray edges \cite{Casandjian2015Fermi}.
\end{enumerate}

\subsection{Modelling Weighted Spatial Template through Lon-Lat Surface Brightness Profiles}
\label{wedge_def}

We make use of the publicly available data\footnote{\url{https://www-glast.stanford.edu/pub_data/1220/GCexcess_Pass8_1704.03910_data/}} from Fermi-LAT \cite{Ackermann2017Fermi}, which includes a residual counts map and its corresponding exposure map. The residual map comprises counts from the central part of the sky map with point sources masked out and consists primarily of the \textit{Fermi} bubbles, inverse Compton (IC) emission, isotropic background and Loop I \cite{Large1962A}. The exposure map consists of the expected counts from sources with respect to the amount of time that they spent in the field of view of the LAT instrument.

\begin{figure}[htbp]
\centering
\subfloat[Extracted longitude profile with Heaviside step functions used in top hat fit.]{
 \includegraphics[height=7.5cm]{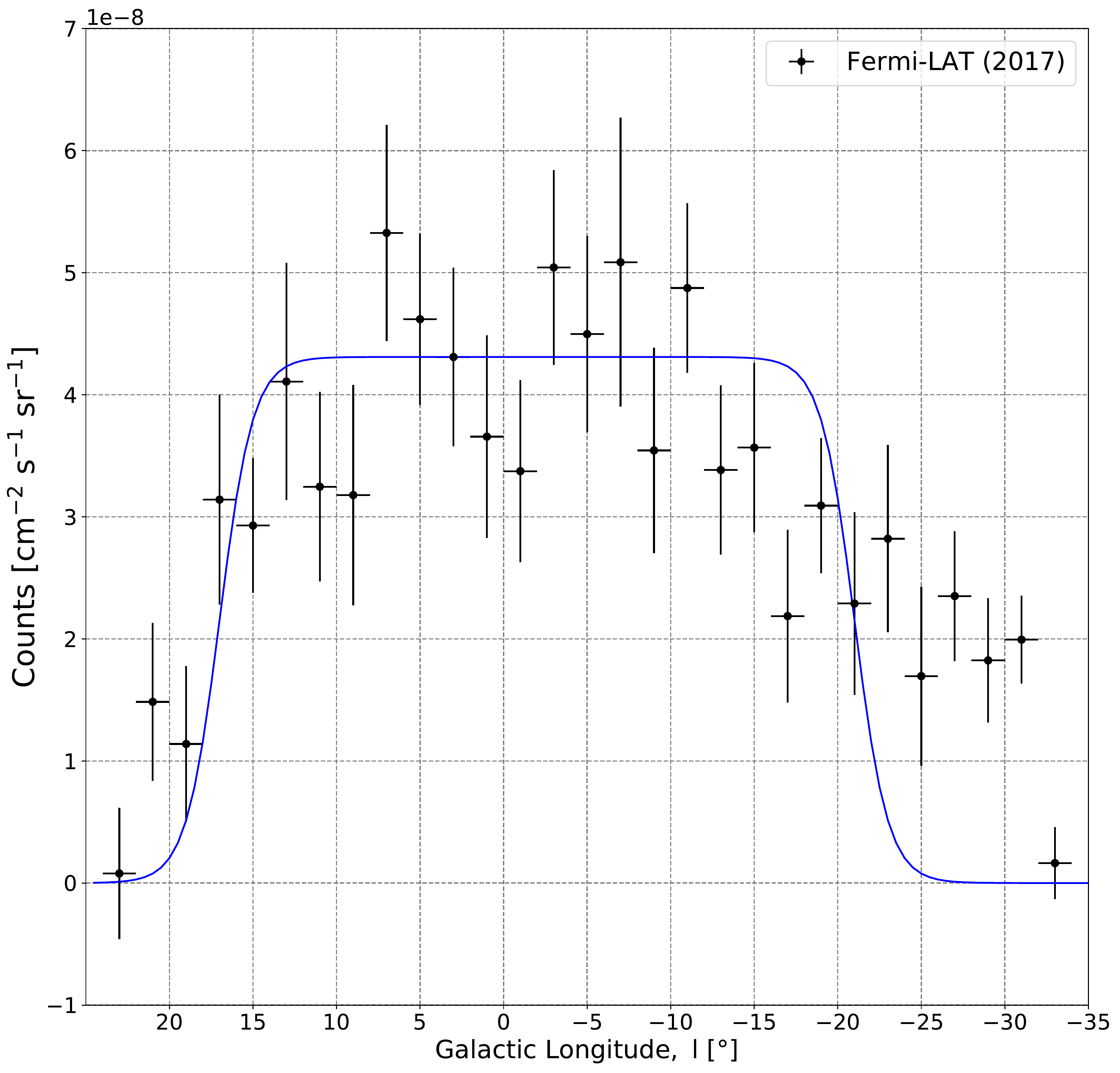}
}
\quad
\subfloat[Extracted latitude surface brightness profile and linear fits applied. A comparison between the wedge, in red, and the NFB, in blue.]{
 \includegraphics[width=6.5cm, height=7.5cm]{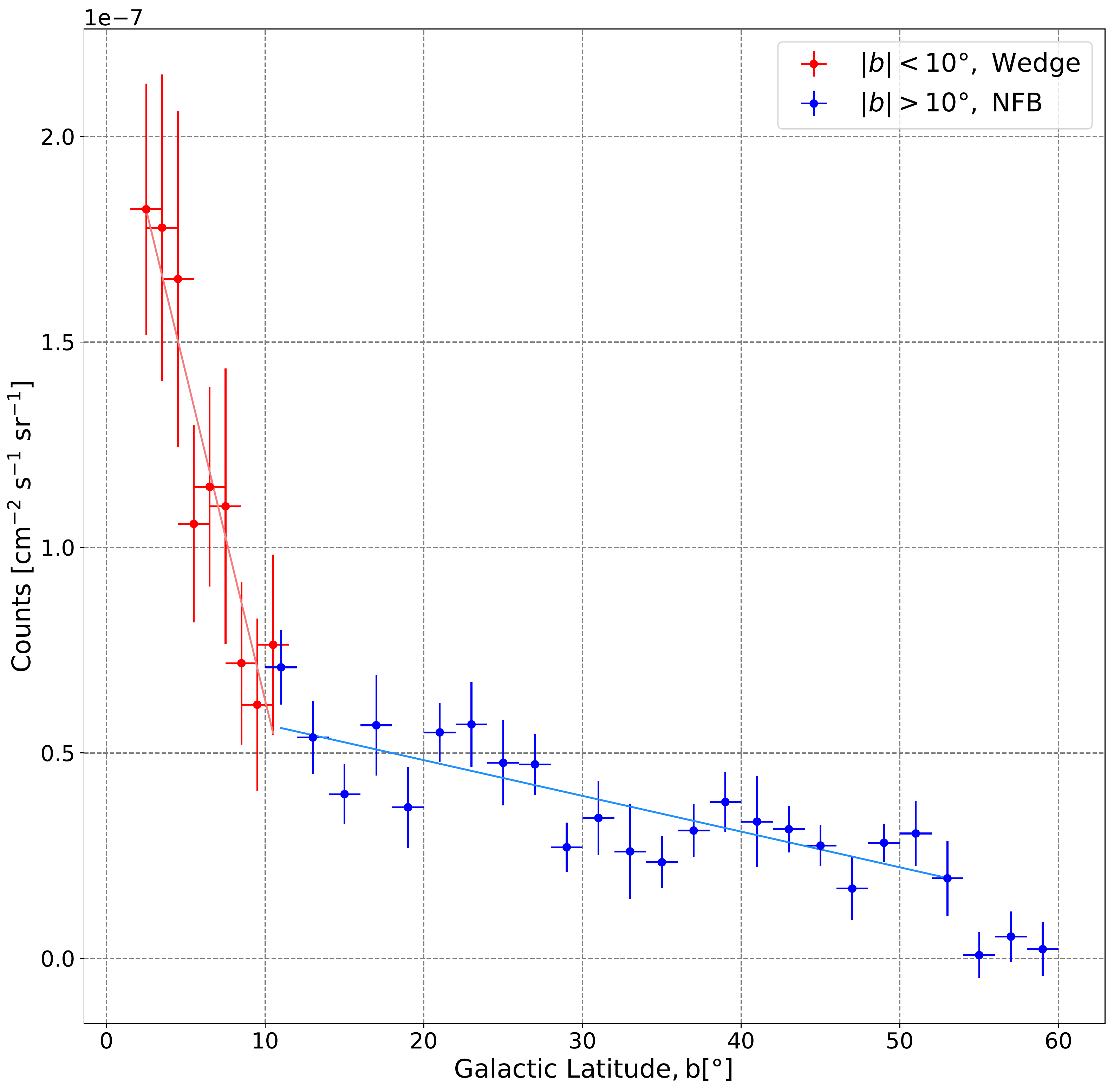}
}
\caption{Longitude and latitude surface brightness profiles for the northern \textit{Fermi} bubble (NFB) and its base (wedge). Data points are exposure corrected counts per solid angle derived from \cite{Ackermann2017Fermi}.}
\label{lonlatprofileextract}
\end{figure}

\medskip
We select longitude and latitude regions on the exposure corrected residual map to extract corresponding surface brightness profiles. For latitude profile where $b>10\degree$, displayed in blue in the right panel of Figure~\ref{lonlatprofileextract}, we essentially sum the exposure corrected counts every $2\degree$ and divide it by the solid angle that it occupies. For $b<10\degree$, using $1\degree$ strips, we extracted the counts for the wedge as well, displayed in red in the right panel of Figure~\ref{lonlatprofileextract}. Since we do not have the original counts map but a modified residual one, we compute the standard error, i.e. root-mean-square of fluctuations about the average value, which we associate with each value extracted from the $2\degree$ (or respectively, $1\degree$ for the wedge) strip. From the data, it is evident that there is brighter emission from the base of the bubble and contrary to the initial result \cite{Ackermann2014Spectrum}, there is a latitude dependence in the surface brightness profile. In principle, a continuous function should describe the surface brightness profile over the entire \textit{Fermi} bubble. However, we choose to work with two distinct linear fits, which can be eventually improved upon if required.

\medskip
We repeat the profile extraction and error computation for longitude. Here, we fit with a top hat function, as depicted in the left panel of Figure~\ref{lonlatprofileextract}, where once more, the fit is not physically motivated but constitutes a simplified starting point which can be refined if required. By convolving the linear fits in latitude and the same top hat fit in longitude for $b<10\degree$ as well as for $b>10\degree$, and then projecting back to the \textit{Fermi} bubble region, we retrieve a \textit{counts} map analogous to the residual map that served as the starting point.

\begin{figure}[htbp]
\centering
\includegraphics[width=0.4\hsize]{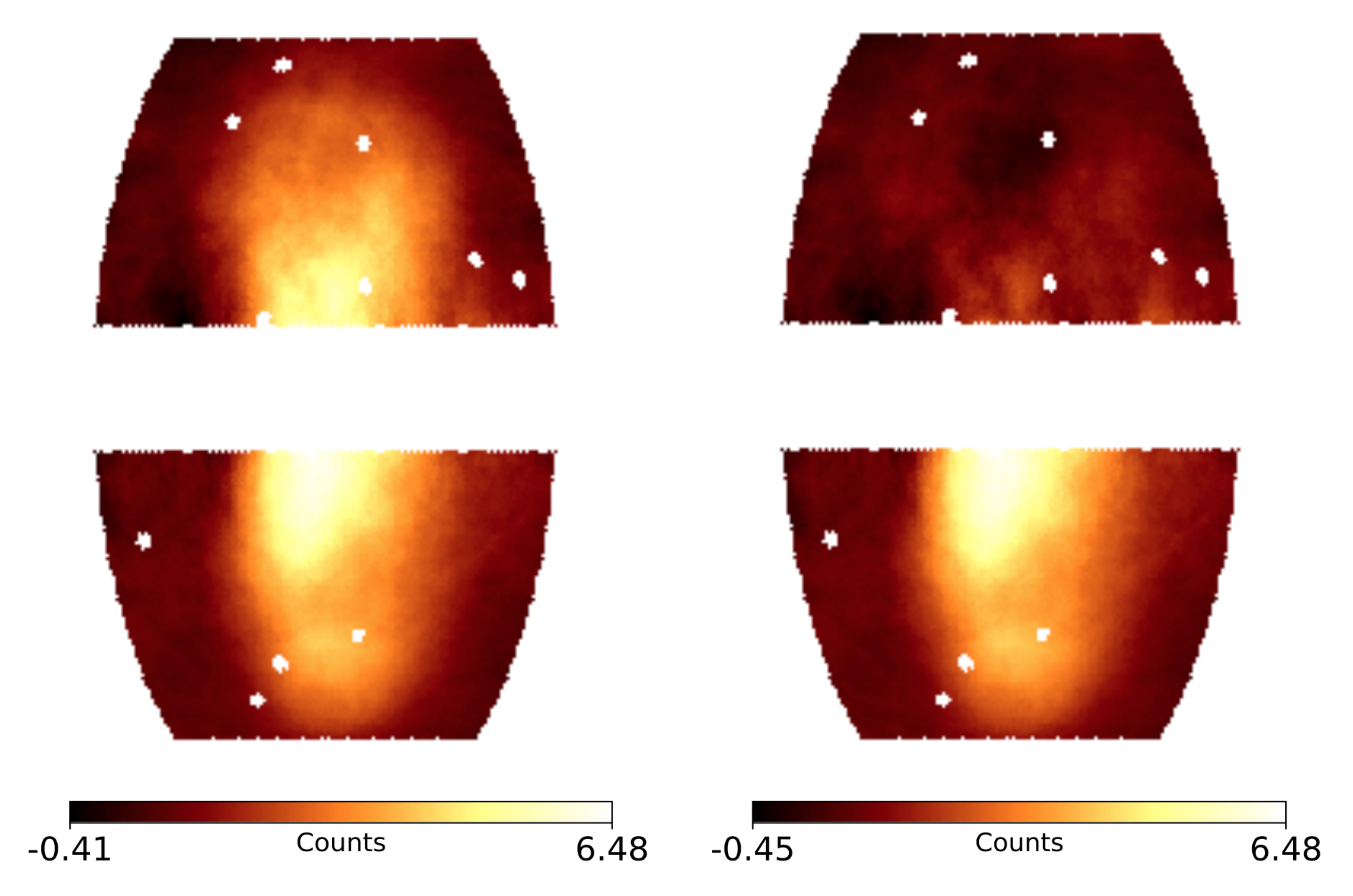}
\caption{Spatial template evaluation. The \textit{left panel} shows the residual counts map from \textit{Fermi}-LAT \cite{Ackermann2017Fermi} smoothed with a $10${\textdegree} integration radius and the \textit{right panel} shows a map with same smoothing after subtraction of the modelled counts from the northern \textit{Fermi} bubble region.}
\label{disappear}
\end{figure}

\medskip
We can verify the validity of the simplistic fits performed on the profiles by generating a difference map. This map is constructed by subtracting the \textit{counts} map from the residual \textit{Fermi}-LAT map. For visualisation purposes, we apply a $10\degree$ smoothing, i.e. we integrate the counts in a $10\degree$ radius for every pixel of both the residual \textit{Fermi}-LAT map and the difference map as depicted in Figure~\ref{disappear}. From the difference map, it is evident that our simplistic fits are adequate as they conveniently describe the northern \textit{Fermi} bubble which could be excised. Now that we have a valid model for expected counts from different parts of the northern \textit{Fermi} bubble, we normalise the sum of counts in the bubble to create a weighted spatial template of the NFB and the wedge.

\section{Model Independent Flux Limits}
\label{Limits}

\begin{figure}[htbp]
\centering
\includegraphics[width=0.5\hsize]{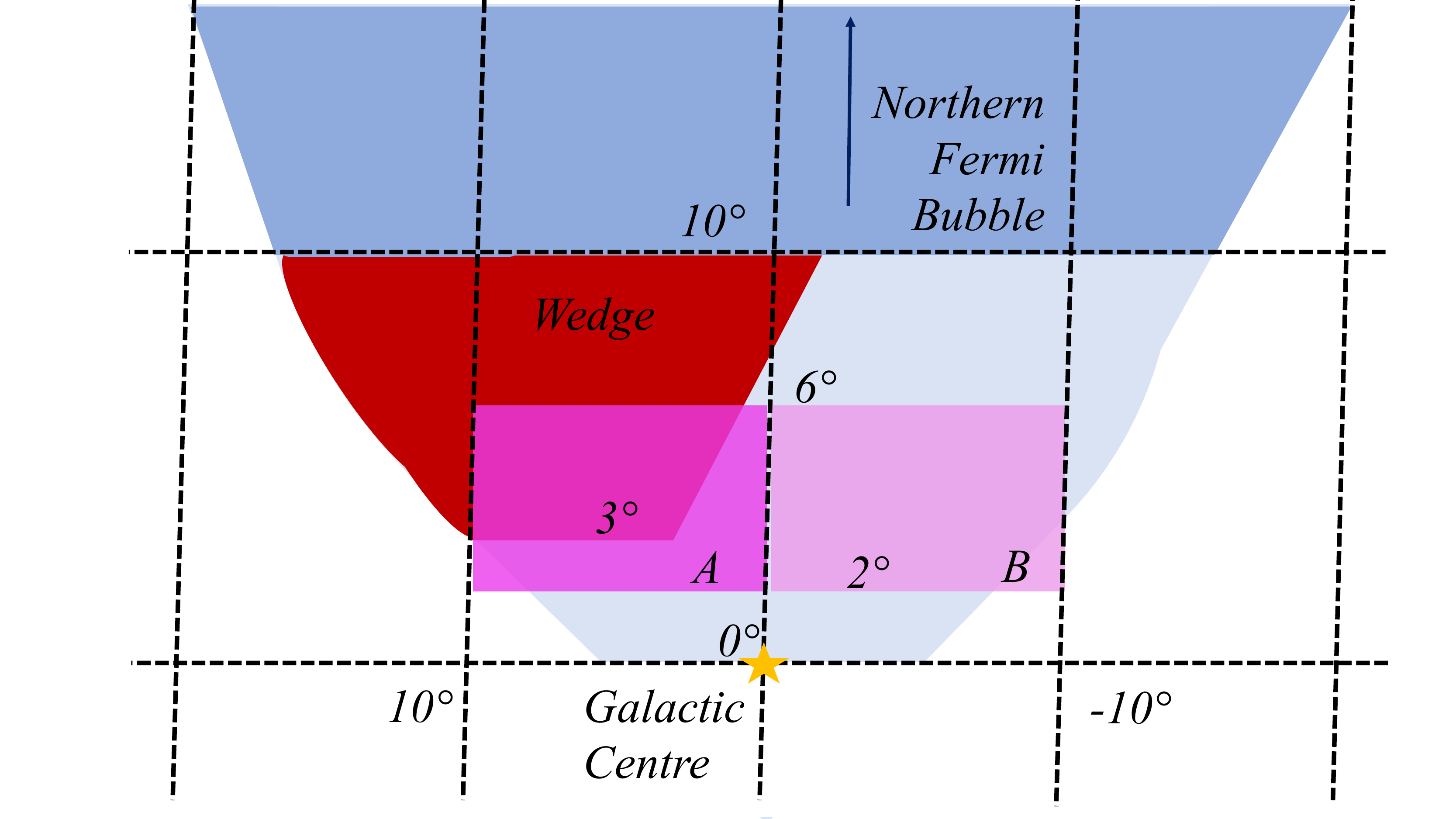}
\caption{Description of regions being compared at the base of the \textit{Fermi} bubble.}
\label{region_des}
\end{figure}

Using the profile likelihood approach described in \cite{Surajbali2020Observing}, we search for an excess signal within the four aforementioned spatial templates. We compare our results for the base of the \textit{Fermi} bubbles to those from \cite{Herold2019Hard} which are also for the base of the \textit{Fermi} bubbles but not exactly the same as the wedge. Figure~\ref{region_des} illustrates the different regions, to scale, whose fluxes are compared. 

\begin{figure}[htbp]
\begin{center}
\includegraphics[height=8cm]{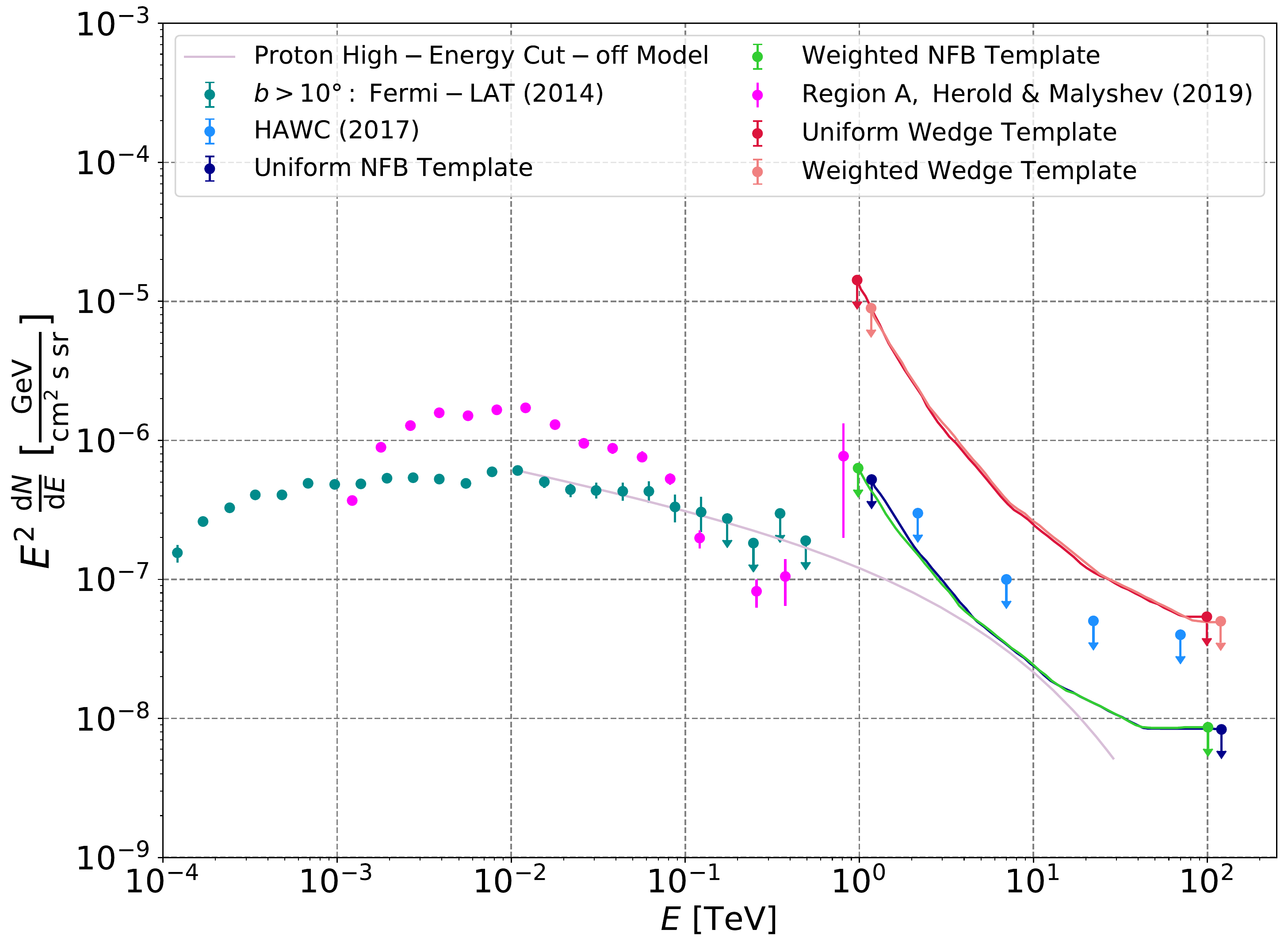}
\end{center}
\caption{Flux upper limit (at 95\% confidence level) comparison between northern \textit{Fermi} bubble (NFB) and wedge, at the base of the bubble. A probable proton high-energy cut-off model, extending to TeV energies, is shown.}
\label{FB_combi_spec}
\end{figure}

\medskip
Figure~\ref{FB_combi_spec} illustrates the observed spectrum and integral flux upper limits of the NFB and the wedge. Although we expected more emission from the wedge, we have a higher flux limit  than for the $b>10\degree$ NFB. This is due to the substantial decline in HAWC sensitivity at low declinations. We fit an exponential cut-off power-law such that it would satisfy both the \textit{Fermi}-LAT observations \cite{Ackermann2014Spectrum} and (barely) our integral flux upper limit. From this fit, the slope $\Gamma$ is $-2.25$ and the cut-off energy occurs at $3.6$ TeV \cite{Surajbali2020Observing}. The relation between $E_{e}$ and $E_{\gamma}$ produced through IC scattering is $E_{e} \approx 11 \sqrt{E_{\gamma}}$ TeV. Hence, in a leptonic scenario for gamma-ray emission, we expect electrons with cut-off energy $\sim 21$ TeV from the northern \textit{Fermi} bubble. For gamma rays produced via hadronic interactions, the energy of the proton and photon energies are related as $E_{p} \approx 23.6\ E_{\gamma}$ TeV \cite{Kappes2007Potential}, with an expected proton cut-off at $\sim 85$ TeV \cite{Surajbali2020Observing}.

\begin{figure}[htbp]
\begin{center}
\includegraphics[height=5cm]{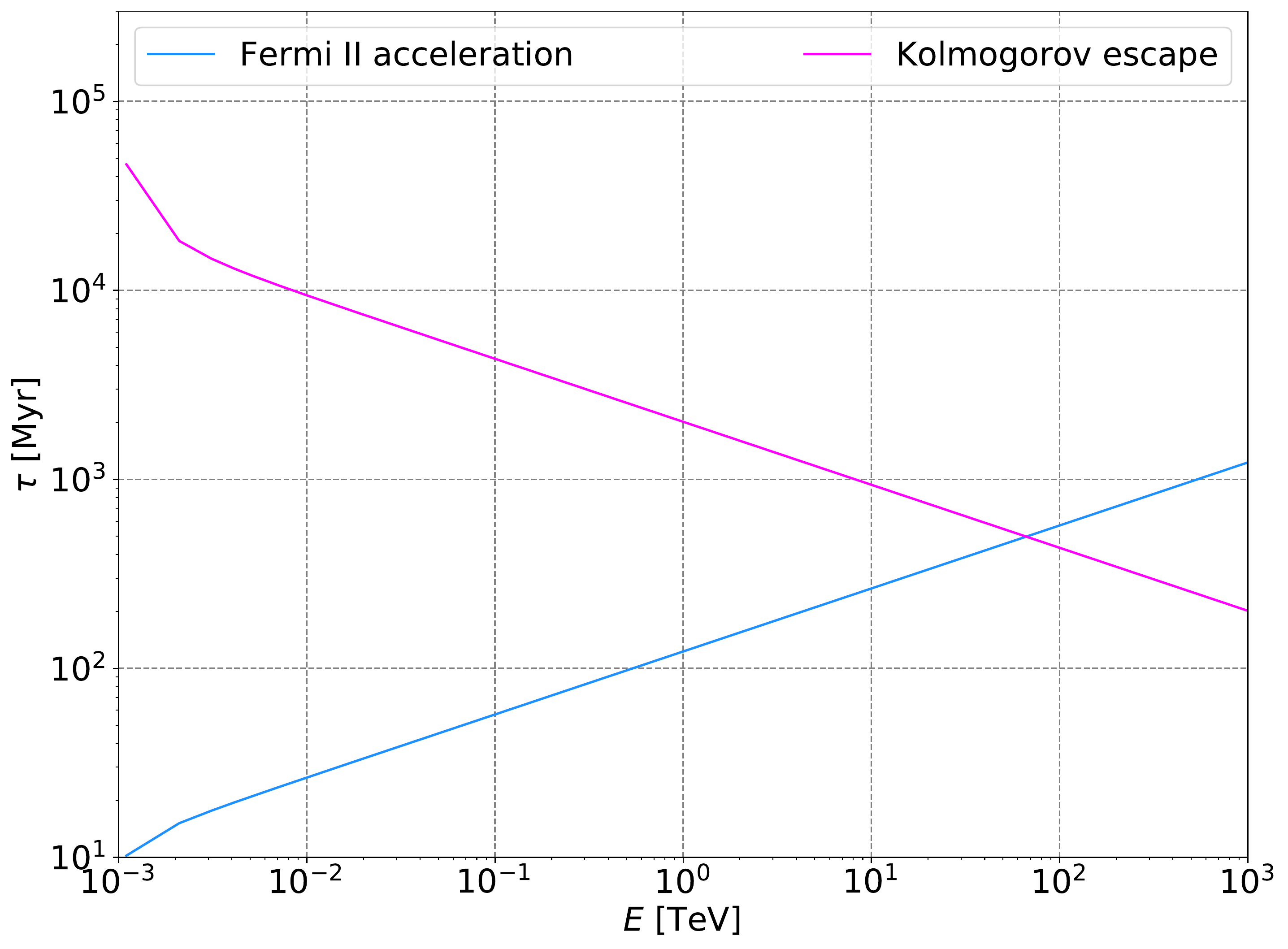}
\end{center}
\caption{Interaction timescales of Fermi second-order acceleration and escape through Kolmogorov diffusion for protons.}
\label{timescaleplot}
\end{figure}

\medskip
To grasp the implications of this high-energy proton cut-off, we depict, in Figure~\ref{timescaleplot}, the acceleration timescale for second-order Fermi acceleration of interacting magnetohydrodynamic (MHD) turbulence with Alfv\'en waves and the escape timescale through diffusion in the Kolmogorov limit. A more detailed description of these timescales can be found in \cite{Schlickeiser2002Cosmic, Surajbali2020Observing}. For this phase-space plot, we varied two parameters: density and magnetic field. We find that a maximum density of $10^{-2}$ cm$^{-3}$ and a minimum magnetic field of $7\ \mu$G satisfy the hadronic upper limit where protons accelerated up to $\sim 85$ TeV are confined and undergo \textit{pp}-interaction while those accelerated above $\sim 85$ TeV escape the bubble \cite{Surajbali2020Observing}. Densities above $10^{-2}$ cm$^{-3}$ would conflict with proton-proton (\textit{pp}) losses. This density and magnetic field are also in agreement with other studies, for instance, \cite{Fujita2014Hadronic} and \cite{ackermann2014fermi}.

\section{Conclusions}
The {\it Fermi} bubbles are few kpc-scale structures that appear to be originating from the central region of our Milky Way galaxy. The physical mechanisms that source the bubbles and produce the observed spectrum are still elusive. Two of the leading gamma-ray production mechanisms involve leptonic or hadronic scenarios, both of which fit the hard gamma-ray spectrum observed and energetics involved but none of them alone can explain all of the observed features and the associated substructures. We performed a template-based search for TeV signals from the northern \textit{Fermi} bubble and just from the base of it. With no significant excess observed, which could be due to the substantial decline in HAWC sensitivity at low declination, we computed integral upper limits at $95\%$ confidence level.

\medskip
Our integral flux upper limits for the northern Fermi bubble are more constraining than the previous limits reported by HAWC while those for the base region provide a more fair comparison to \textit{Fermi}-LAT data points from the base of the bubble. For the northern Fermi bubble, we present a hadronic model with proton cut-off energy at $85$ TeV that fits the \textit{Fermi}-LAT data and conforms to our flux limits. This fit is a power-law with index $-2.25$ and cut-off at $3.6$ TeV \cite{Surajbali2020Observing}. Using the high-energy proton cut-off, we further constrain the density and magnetic field of the bubble to be $\sim 10^{-2}$ cm$^{-3}$ and $\sim 7\ \mu$G, respectively \cite{Surajbali2020Observing}.

\section*{Acknowledgements}
We acknowledge the support from: the US National Science Foundation (NSF); the US Department of Energy Office of High-Energy Physics; the Laboratory Directed Research and Development (LDRD) program of Los Alamos National Laboratory; Consejo Nacional de Ciencia y Tecnolog\'ia (CONACyT), M\'exico, grants 271051, 232656, 260378, 179588, 254964, 258865, 243290, 132197, A1-S-46288, A1-S-22784, c\'atedras 873, 1563, 341, 323, Red HAWC, M\'exico; DGAPA-UNAM grants IG101320, IN111716-3, IN111419, IA102019, IN110621, IN110521; VIEP-BUAP; PIFI 2012, 2013, PROFOCIE 2014, 2015; the University of Wisconsin Alumni Research Foundation; the Institute of Geophysics, Planetary Physics, and Signatures at Los Alamos National Laboratory; Polish Science Centre grant, DEC-2017/27/B/ST9/02272; Coordinaci\'on de la Investigaci\'on Cient\'ifica de la Universidad Michoacana; Royal Society - Newton Advanced Fellowship 180385; Generalitat Valenciana, grant CIDEGENT/2018/034; Chulalongkorn University’s CUniverse (CUAASC) grant; Coordinaci\'on General Acad\'emica e Innovaci\'on (CGAI-UdeG), PRODEP-SEP UDG-CA-499; Institute of Cosmic Ray Research (ICRR), University of Tokyo, H.F. acknowledges support by NASA under award number 80GSFC21M0002. We also acknowledge the significant contributions over many years of Stefan Westerhoff, Gaurang Yodh and Arnulfo Zepeda Dominguez, all deceased members of the HAWC collaboration. Thanks to Scott Delay, Luciano D\'iaz and Eduardo Murrieta for technical support.

\bibliographystyle{JHEP}
\bibliography{ref}

\clearpage
\section*{Full Authors List: \Coll\ Collaboration}
%

\scriptsize
\noindent
A.U. Abeysekara$^{48}$,
A. Albert$^{21}$,
R. Alfaro$^{14}$,
C. Alvarez$^{41}$,
J.D. Álvarez$^{40}$,
J.R. Angeles Camacho$^{14}$,
J.C. Arteaga-Velázquez$^{40}$,
K. P. Arunbabu$^{17}$,
D. Avila Rojas$^{14}$,
H.A. Ayala Solares$^{28}$
% ...

\noindent
% ...
$^{14}$Instituto de Física, Universidad Nacional Autónoma de México, Ciudad de México, México,
% ...
$^{17}$Instituto de Geofísica, Universidad Nacional Autónoma de México, Ciudad de México, México,
% ...
$^{21}$Physics Division, Los Alamos National Laboratory, Los Alamos, NM, USA,
% ...
$^{28}$Department of Physics, Pennsylvania State University, University Park, PA, USA,
% ...
$^{40}$Instituto de Física y Matemáticas, Universidad Michoacana de San Nicolás de Hidalgo, Morelia, Michoacán, México,
$^{41}$FCFM-MCTP, Universidad Autónoma de Chiapas, Tuxtla Gutiérrez, Chiapas, México,
% ...
$^{48}$Department of Physics and Astronomy, University of Utah, Salt Lake City, UT, USA,
% ...

\end{document}